# Probing phonon chirality and circular lattice motion with symmetry-selective nonlinear optical spectroscopy

**Authors:** Yuhan Wang[1], Yuxuan Wei[1], Li Huang[1] and Chuanshan Tian[1,2,*]

**Affiliations:**

[1]Department of Physics, State Key Laboratory of Surface Physics and Key Laboratory of Micro- and Nano-Photonic Structure (MOE), Fudan University, Shanghai 200433, China

[2]Institute of Quantum Science and Technology, Yanbian University, Yanji, Jilin 133002, China

*To whom correspondence should be addressed: cstian@fudan.edu.cn

## Abstract

Truly chiral phonons are lattice eigenmodes that combine broken mirror symmetry with circular atomic motion. They can mediate angular-momentum-selective interactions in quantum materials, yet directly resolving both their chirality and underlying circular motion remains challenging, especially in high-symmetry crystals. Here we show that symmetry-selective terahertz difference-frequency spectroscopy provides a phase- and polarization-resolved route to identifying truly chiral phonons on a tabletop experiment. Using $\alpha$-quartz as a benchmark, we validate this approach by resolving phonon chirality via chiral-sensitive $\chi^{(2)}_{ijk}$ tensor elements ($i \neq j \neq k$), while vector-field detection directly reveals a time-dependent polarization rotation arising from circular ionic motion and thus nonzero angular momentum. Applying the same protocol to tetragonal $\alpha$-TeO$_2$, we isolate chiral $E$-mode resonances below 5 THz and directly verify their circular lattice motion, thereby resolving a symmetry-imposed ambiguity in chiral-phonon identification in fourfold-symmetric crystals. Our results establish symmetry-selective nonlinear terahertz spectroscopy as a general route to identify truly chiral phonons in condensed matter systems.

# Introduction

Chirality arises from the breaking of improper symmetries that manifests across modern science, from parity violation in particle physics to biological homochirality.[1-6] In crystalline solids, phonons, long treated as achiral vibrations, can execute circular or elliptical ionic motion and thereby carry angular momentum (AM)[7,8]. In enantiomorphic environments, chirality lifts the degeneracy between opposite phonon helicities,[9] resulting in helicity-dependent phonon couplings to light, electrons and other collective excitations. Thus, chiral lattice motion has emerged as an active degree of freedom in quantum and functional materials, with important consequences for angular-momentum transfer and transport phenomena, including orbital Seebeck effect[10], thermal Hall effect[11,12], Einstein-de Haas effect[13].

Despite this progress, a central conceptual and experimental challenge lies in disentangling chirality from real-space circular motion. These two notions are, in general, independent[7]: a phonon can be chiral while remaining purely linearly polarized without rotational motion[14], while circular or elliptical ionic trajectories can also occur in non-chiral systems[15]. Real-space angular momentum (AM) is associated with actual rotational motion of atoms. By contrast, pseudo-angular momentum (PAM) arises from symmetry-imposed phase factors of vibrational wavefunctions under discrete operations and reflects symmetry properties rather than physical rotation[9]. This fundamental distinction implies that neither chirality nor circular motion alone is sufficient to establish a truly chiral phonon; instead, one must simultaneously resolve both symmetry-defined chirality and real-space rotational dynamics. Consequently, experimentally identifying such modes remains highly nontrivial.

Existing experimental approaches have provided important insights but face key limitations. Circularly polarized resonant inelastic X-ray scattering (RIXS) grants access to chiral-phonon dispersion at finite momenta, yet requires synchrotron-based instrumentation and currently lacks the energy resolution needed to resolve individual low-energy phonon branches[16]. Terahertz circular dichroism (TCD)[17] can probe low-energy chiral vibrations in isotropic molecular assemblies, but in crystal its sensitivity is often compromised by anisotropy-induced linear dichroism and birefringence backgrounds, while intrinsic optical activity in the terahertz regime is intrinsically small[17]. Circularly polarized Raman scattering (CP-Raman)[9], combined with first-principles calculations, has enabled the identification of truly chiral phonons and clarified the role of PAM selection rules. However, Raman processes probe PAM rather than real-space AM, and become ineffective in many systems like fourfold-symmetric (tetragonal) crystals, where symmetry-enforced degeneracies lead to identical responses for different circular polarizations (see details in SI) [18]. This limitation is particularly significant because tetragonal and near-tetragonal motifs are pervasive in technologically important materials classes, including complex oxides, ferroics and unconventional superconductors[12,15,19,20].

Beyond these technique-specific constraints, a more fundamental limitation is that chiral signals in most of these spectroscopies originate from electric-dipole-forbidden processes and are therefore intrinsically weak [21]. This motivates the search for electric-dipole-allowed probes of chiral lattice dynamics. Second-order nonlinear optical techniques, such as second-harmonic generation (SHG)[22] and sum-frequency generation (SFG)[23,24], are governed by the rank-3 susceptibility tensor $\chi^{(2)}$, which matches the 3D nature of chirality[25,26]. Importantly, these processes are electric-dipole-allowed and can therefore yield substantially stronger chiral signals. However, conventional SHG and SFG primarily probe chirality and do not directly resolve the underlying atomic motion of chiral phonons, in particular the real-space circular lattice dynamics associated with AM.

Here we introduce terahertz difference frequency spectroscopy (THz-DFS)[27] as a symmetry-sensitive probe that directly accesses both phonon chirality and coherent circular lattice motion. THz-DFS exploits the second-order susceptibility $\chi^{(2)}$ to access enantiomer-odd responses via electric-dipole-allowed interactions, providing a sensitive and unambiguous fingerprint of phonon chirality. Crucially, field-resolved detection of the emitted terahertz radiation retrieves its full vector character, enabling direct visualization of time-dependent polarization rotation associated with circular ionic motion. We first validate this approach in α-quartz[10,16,28], where chiral-sensitive $\chi^{(2)}$ tensor elements (i ≠ j ≠ k) reveal phonon chirality, while vector-field measurements directly capture the corresponding circular dynamics. We then apply the same protocol to tetragonal α-TeO$_2$, isolating chiral E-mode resonances below 5 THz and resolving the long-standing ambiguity in chiral-phonon identification in fourfold-symmetric systems. Our results establish symmetry-selective nonlinear terahertz spectroscopy as a general tabletop route for identifying truly chiral phonons and directly visualizing their coherent lattice motion in quantum materials.

## Probing Chiral phonons in Quartz: handedness and circular lattice motion

We choose enantiomorphic α-quartz (SiO$_2$) as a benchmark system. Its phonon chirality is well established, and its low-frequency phonon branches have been extensively studied[10,16,28]. Near the Γ point, the doubly degenerate $E$ ($\Gamma_3$) modes of α-quartz simultaneously carry well-defined pseudo-angular momentum (PAM = ±1) and finite real angular momentum (AM ≠ 0), while being both infrared- and Raman-active, as summarized in Table 1. This coexistence makes α-quartz a unique platform where chirality, PAM, and AM are all present, allowing us to validate whether THz-DFS can (i) identify phonon chirality and (ii) directly resolve real-space circular lattice motion.

**Table 1  Phonon Modes in D$_3$(32) Point Group**

| Phonon mode | IR | Raman | $\chi^{(2)}$ | PAM | AM |
|---|---|---|---|---|---|
| A$_1$ | 0 | A$_1$ | 0 | 0 | 0 |
| A$_2$ | z | 0 | 0 | 0 | 0 |
| E | x, y | E | $\chi^{(2)}_E$ | ±1 | ≠0 |

To first identify phonon chirality, we measure the time-domain terahertz waveform $E_{THz}(t)$ emitted via chiral-sensitive $\chi^{(2)}_{ijk}$ tensor elements (i ≠ j ≠ k) of the DFS process (Methods). These tensor components are nonzero only when the system is chiral, and are odd under mirror reflection, (i.e. reverse sign between opposite enantiomers), thereby providing a symmetry-protected probe of chirality. Under excitation condition in S-, P-, P-polarization (S out, P/P in) combination[24], the chiral $\chi^{(2)}_{yxz}$ and $\chi^{(2)}_{yzx}$ channels are selectively excessed (Fig. 1a and Methods). Figure 1c-d compares the emitted THz fields from *z*-cut left- and right-handed *α*-quartz (Fig. 1b) under identical excitation conditions. Both enantiomers exhibit a pronounced resonance at 3.7 THz (128 cm$^{-1}$), corresponding to the lowest-frequency chiral *E*-mode phonon (Fig. 1e). Crucially, the time-domain waveform undergoes a global sign reversal between the two enantiomers, while keeping the resonance frequency and lineshape invariant between the two enantiomers (Fig. 1d). This enantiomer-locked phase reversal provides a stringent and internally self-referenced fingerprint of phonon chirality[16]. Over a broader frequency window (Extended Data Fig. 1), multiple resonances are resolved that map directly onto all symmetry-allowed chiral optical phonon branches of *α*-quartz shown in Fig. 1e, whereas non-chiral modes inactive in the chiral-sensitive $\chi^{(2)}_{yxz}$ channel remain below detection. Notably, as an electric-dipole-allowed process, the chiral DFS is intrinsically insensitive to residual birefringence and achiral backgrounds in crystals.

Having established chirality sensitivity, we next examine whether these phonons carry real angular momentum (AM), which cannot be inferred from chirality or PAM alone. We therefore use two oppositely circularly polarized pumps ($\omega_1$ with $\sigma_1$ and $\omega_2$ with $\sigma_2$) to selectively drive angular-momentum-carrying *E* phonon mode,[9,16] and detect the full vector terahertz field emission at $\Omega = \omega_1 - \omega_2$ (Fig. 2a and Methods). In *α*-quartz, the transfer of PAM from light to the lattice governs the selection rule[9] $m = \sigma_1 - \sigma_2 - 3p$ ($p = 0, \pm 1$), where $\sigma_{1,2} = \pm 1$ denote the helicities of the incident photons and *m* is the PAM of the excited phonons (Fig. 2b). While the PAM selection rule determines which phonon modes are excited, it does not guarantee the phonon has finite AM. If the excited phonons do carry AM, their circular ionic motion generates a rotating dipole, that emit a terahertz field whose polarization vector rotates in time.

As shown in Fig. 2c, the emitted terahertz field from left-handed α-quartz exhibits a clear temporal polarization rotation, providing direct evidence of circular lattice motion. Reversing the helicities of the two pumps reverses the rotation sense of the emitted field (Fig. 2d), and the same behaviour is observed for right-handed quartz (Fig. S5 in SI). Notably, the rotation sense is determined by the helicity configuration of the incident pumps, that is, by the PAM transferred to the lattice, and is independent of the chirality of the crystal. Taken together with phase reversal of the chiral-sensitive $\chi^{(2)}_{yxz}$ channel described above, these results demonstrate that the 128 cm$^{-1}$ E-mode in α-quartz simultaneously exhibits (i) chirality and (ii) finite real-space angular momentum, thereby fulfilling the criteria for a truly chiral phonon.

To explicitly disentangle phonon chirality from angular momentum, we introduce crystalline sucrose as a control system. Although sucrose exhibits a finite enantiomer-odd $\chi^{(2)}_{yxz}$ response indicative of phonon chirality, the vibrational modes at 1.4 THz and 1.8 THz are essentially linear and therefore carry zero AM (Extended Data Fig. 2). Conversely, even if circular ionic motion (finite AM) is present, the $\chi^{(2)}_{ijk}$ ($i \neq j \neq k$) channel vanishes if the phonon is non-chiral (see details in method). These observations establish a stringent criterion: truly chiral phonons require the simultaneous presence of chirality and real-space angular momentum. THz-DFS uniquely enables both conditions to be verified within a single experimental framework.

**Chirality and Circular lattice motion in tetragonal α-TeO₂**

We then apply the same protocol to tetragonal α-TeO$_2$ (paratellurite), a structurally chiral crystal belonging to the enantiomorphic space groups P4$_3$2$_1$2 (left-handed) or P4$_1$2$_1$2 (right-handed) (Fig. 3a). In this system, the presence of fourfold (C4) symmetry renders chiral discrimination challenging for conventional CP-Raman probes[18]. The distorted TeO$_4$ units form a helical arrangement along the c axis, and the primitive cell contains 12 atoms, giving rise to a total of 33 optical phonon modes with the Γ-point decomposition

$$\Gamma_{\text{opt, TeO2}} = 4A_1 + 4A_2 + 5B_1 + 4B_2 + 8E$$

Among these, only the doubly degenerate $E$ modes are both infrared- and Raman-active; they are therefore the only optical phonons that can contribute to the present DFS channel, whereas the $A_1$, $B_1$, and $B_2$ modes are infrared-inactive and the $A_2$ modes are Raman-inactive, and therefore yield a null $\chi^{(2)}$ response as shown in Table 2.

Table 2  Phonon Modes in $D_4(422)$ Point Group

| Phonon Mode | IR | Raman | $\chi^{(2)}$ | PAM | AM |
|---|---|---|---|---|---|
| $A_1$ | 0 | $A_1$ | 0 | 0 | 0 |
| $A_2$ | z | 0 | 0 | 0 | 0 |
| $B_1$ | 0 | $B_1$ | 0 | 2 | 0 |
| $B_2$ | 0 | $B_2$ | 0 | 2 | 0 |
| $E$ | x, y | $E$ | $\chi^{(2)}_E$ | ±1 | ≠0 |

We first probe phonon chirality using the chiral-sensitive $\chi^{(2)}_{yxz}$ response. Fig. 3b presents the THz-DFS spectra of $\alpha$-TeO$_2$ measured in the chiral $\chi^{(2)}_{yxz}$ geometry. Within our measurement bandwidth, two prominent resonances appear at 3.6 THz and 5.3 THz, which are assigned to $E$-symmetry phonons based on calculation[29] (Fig. 3c). By contrast, those achiral modes of A or B symmetry (marked in Fig. 3b) remain silent. This selectivity demonstrates that the chiral-sensitive $\chi^{(2)}_{yxz}$ channel provides a direct and background-resistant fingerprint of phonon chirality even in a tetragonal system.

We next examine whether these modes also carry AM. Using circularly polarized difference-frequency excitation, we control the PAM transferred from the optical fields to the lattice. In a fourfold-symmetric crystal, this transfer follows the selection rule $m = \sigma_1 - \sigma_2 - 4p$ ($p = 0, \pm 1$). To access AM-carrying modes, we employ an oblique-incidence geometry, in which $m = \pm 1$ and $m = 2$ modes can be simultaneously excited. Because the $m = 2$ mode (AM = 0, see Extended Data Fig. 3) is infrared-inactive, its DFS response vanishes and the detected emission contains only the $m = \pm 1$ phonon contributions. As shown in Fig. 3d, the emitted THz field from α-TeO$_2$ exhibits a clear time-dependent polarization rotation with nearly circular character, providing direct evidence of coherent circular lattice motion. Reversing the helicities of the two incident fields reverses the rotation sense of the emitted polarization (Extended Data Fig. 4), demonstrating control of the phonon AM. Notably, for the same excitation helicity configuration, the rotation sense of the emitted terahertz field is opposite to that observed in α-quartz, reflecting the different PAM selection rule in the two materials.

Together with the $\chi^{(2)}_{yxz}$-based chirality criterion, these results demonstrate that the observed *E*-mode phonons in α-TeO$_2$ simultaneously exhibit (i) chirality and (ii) finite AM, and therefore constitute truly chiral phonons. From the temporal envelope of the emitted terahertz field, we extract a room-temperature coherence lifetime of 0.87 ps, shorter than that in α-quartz (2.5 ps), primarily due to enhanced anharmonicity arising from the stereochemically active lone-pair electrons of Te$^{4+}$, which increase phonon-phonon scattering[30].

## Conclusion

In summary, we establish terahertz difference-frequency time-domain spectroscopy as a symmetry-selective, phase- and polarization-resolved route to truly chiral phonons. The key advance is that a single optical platform unifies two complementary observables: an enantiomer-odd phase criterion that provides a stringent fingerprint of phonon chirality, and a polarization-resolved vector-field readout that directly reveals real-space circular lattice motion and thus angular momentum (AM). By combining these two criteria, the method distinguishes chirality from angular-momentum-carrying phonons and resolves a long-standing challenge in identifying chiral phonons in high-symmetry crystals. Because the response is mediated by electric-dipole-allowed lattice excitations, the approach further offers high sensitivity in the terahertz regime. More broadly, it establishes a general framework for driving and quantifying helical lattice motion, and opens new opportunities for exploring how lattice angular momentum couples to electronic, magnetic and topological excitations in quantum materials.

# Figures and captions

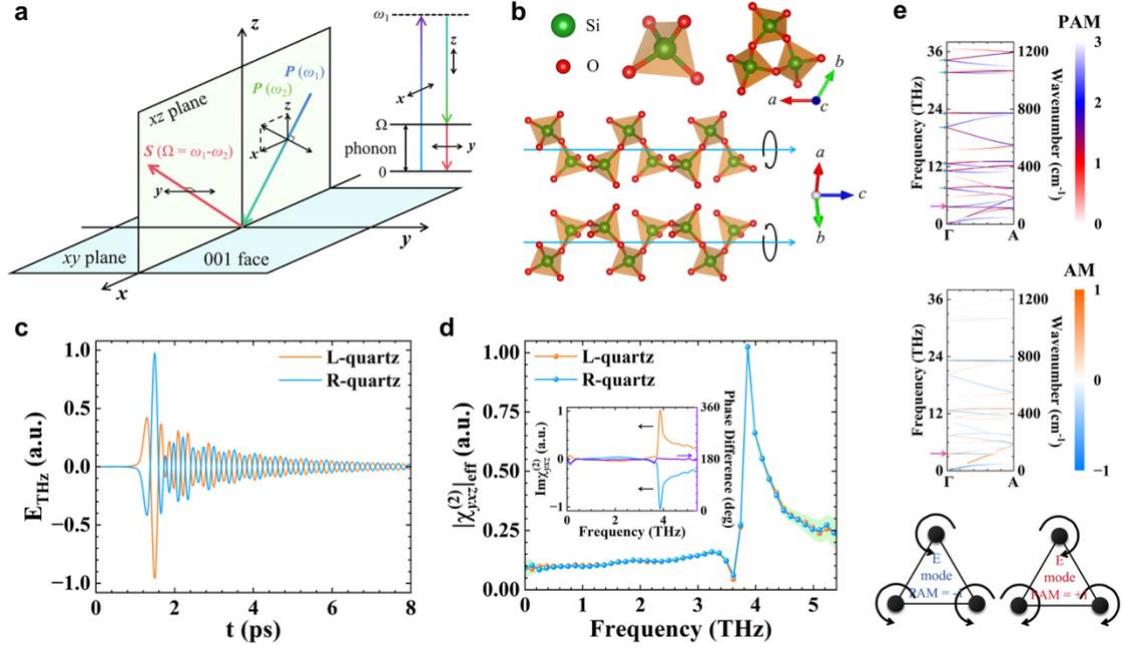

**Fig. 1: Chirality-sensitive THz-DFS detection of chiral phonons in α-quartz. a,** Schematic of the chiral terahertz difference-frequency spectroscopy (Chiral THz-DFS) geometry. The emitted terahertz field is detected in a chirality-sensitive $\chi^{(2)}_{yxz}$ channel. **b,** Crystal structure of α-quartz ($SiO_2$), showing its enantiomorphic lattice[31]. **c,** Emitted terahertz waveforms from opposite enantiomers exhibit a global sign reversal. **d,** The nonlinear susceptibility and its imaginary part (insert) shows a π-phase difference (insert) between enantiomers, while the resonance frequency remains unchanged. **e,** Calculated[32] phonon dispersion[33] of α-quartz along the Γ-A direction, with color indicating pseudo-angular momentum (PAM, top) and real angular momentum (AM, middle). Only doubly degenerate E modes (red and green arrow pointing) carry nonzero PAM and finite AM near Γ point. Bottom: schematic illustration of circular motion for E modes with opposite angular momentum.

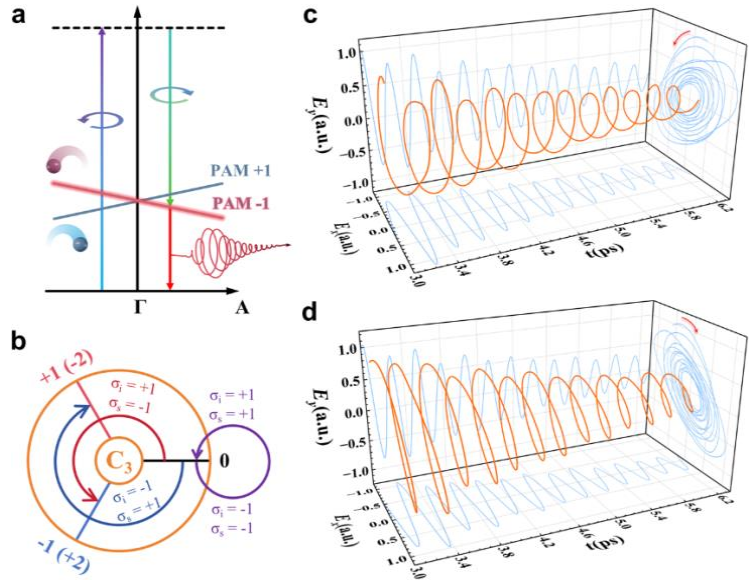

**Fig. 2: Angular-momentum transfer and circular lattice motion in *α*-quartz. a**, Circularly polarized difference-frequency excitation transfers pseudo-angular momentum (PAM) from light to lattice vibrations, selectively driving phonons with PAM = ±1. The resulting circular motion (non-zero AM) radiates a terahertz field whose polarization rotates in time. **b,** Selection rules under $C_3$ symmetry link the helicities of the incident fields to the PAM of the excited phonons. **c,** The emitted terahertz field exhibits a rotating polarization in time, directly visualizing coherent circular lattice motion. **d,** Reversing the pump helicities inverts the rotation sense of the terahertz polarization, evidencing controlled reversal of phonon angular momentum. Full time-domain spectra is shown in Fig. S4, frequency spectra is shown in Fig. S5.

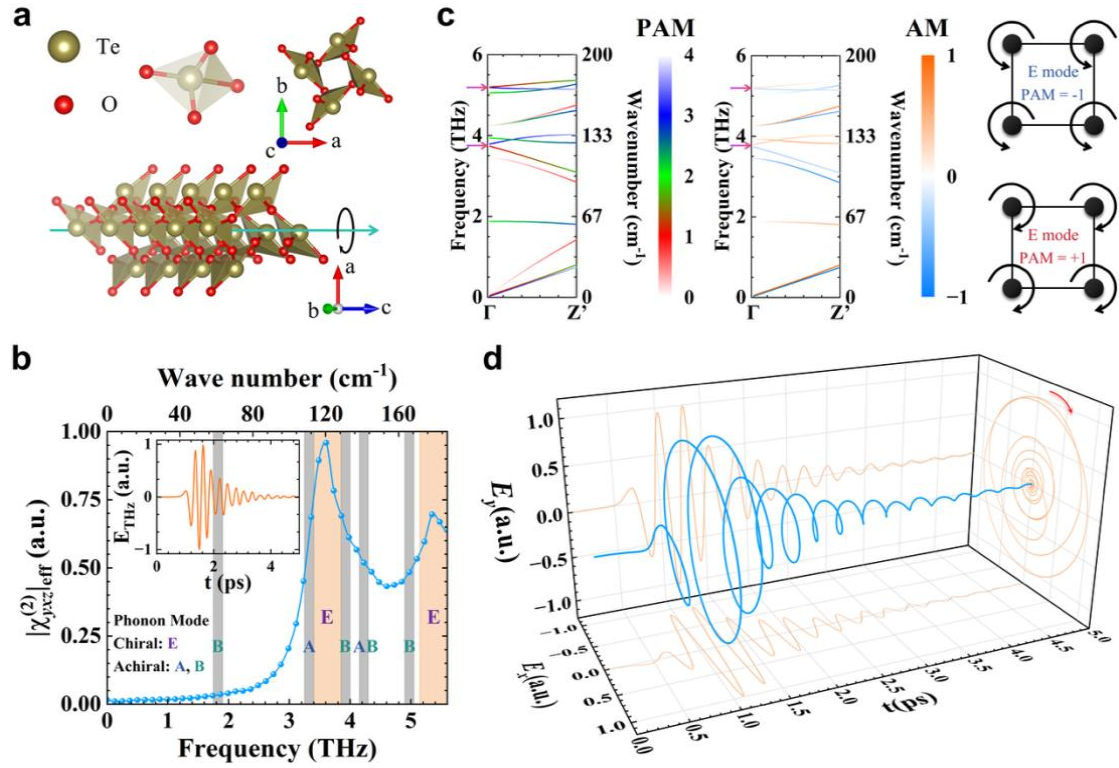

**Fig. 3: Symmetry-selective identification of chiral phonons in tetragonal α-TeO₂. a**, Crystal structure of α-TeO$_2$, showing the helical arrangement of Te-O tetrahedron in a tetragonal (C$_4$) lattice. **b**, Time (inset) and frequency domain THz-DFS in a chiral-sensitive channel. Only E-symmetry phonons appear, while A- and B-type (achiral) modes remain silent. **c**, Calculated phonon dispersion with PAM and AM. Only doubly degenerate E modes carry PAM = ±1 and finite AM. **d,** The emitted terahertz field exhibits a rotating polarization in time, demonstrating coherent circular lattice motion.

## Methods

**Terahertz difference-frequency spectroscopy**

Terahertz difference-frequency spectroscopy (THz-DFS) was performed using a Yb:KGW laser (PHAROS, Light Conversion) operating at a central wavelength of 1030 nm, delivering 190 fs pulses at a repetition rate of 100 kHz. The pulses were externally compressed to 26 fs using a home-built compressor[34]. The output beam was split by a 9:1 beam splitter, with 90% of the pulse energy directed to the sample to generate terahertz radiation via second-order difference-frequency processes, and the remaining 10% used as a probe for detection.

The emitted terahertz electric field was measured in the time domain by electro-optic sampling, enabling phase-resolved detection of both amplitude and polarization of the field. This allows direct reconstruction of the vectorial terahertz waveform associated with the lattice response. The measured difference-frequency spectra were normalized using GaP as a reference sample. In the frequency range below 5 THz, GaP exhibits an essentially dispersionless second-order nonlinear susceptibility, ensuring reliable normalization (see SI and Ref.[27] for further details on the experimental setup and data processing).

**Chiral Terahertz Difference Frequency Spectroscopy**

Under reflection geometry, the THz-DFS field at frequency $\Omega = \omega_1 - \omega_2$ can be expressed as

$$\boldsymbol{E}_{\text{THz}} \propto \vec{\boldsymbol{L}}(\Omega) \cdot \boldsymbol{\chi}^{(2)}(\Omega; \omega_1, -\omega_2) : [\vec{\boldsymbol{L}}_1(\omega_1) \cdot \boldsymbol{E}_1(\omega_1) \otimes \vec{\boldsymbol{L}}_2(\omega_2) \cdot \boldsymbol{E}_2^*(\omega_2)] / \Delta k \qquad (1)$$

where $k_i$, $\boldsymbol{E}_i(\omega_i)$, and $\vec{\boldsymbol{L}}_i(\omega_i)$ denote the wave vector, electric field, and Fresnel transmission tensor at frequency $\omega_i$, respectively. Here, $\chi^{(2)}$ represents the second-order nonlinear susceptibility, and $\Delta k = k_{\omega_1} - k_{\omega_2} + k_\Omega$ is the phase mismatch. In reflection geometry, $\Delta k$ can be approximated as

$$\Delta k \approx 4\pi n_\Omega / \lambda_\Omega \qquad (2)$$

where $n_\Omega$ and $\lambda_\Omega$ are the refractive index and wavelength of the sample at the terahertz frequency.

For the SPP configuration at an incident angle of 45°, the emitted THz field can be expressed as

$$\boldsymbol{E}_{\text{THz, s}} \propto \chi^{(2)}_{spp} \boldsymbol{E}_p^2 = 2\chi^{(2)}_{yxx} L_{yy}(\Omega) L_{xx}(\omega_1) L_{xx}(\omega_2) \boldsymbol{E}_p^2 + 2\chi^{(2)}_{yzz} L_{yy}(\Omega) L_{zz}(\omega_1) L_{zz}(\omega_2)$$

$$\boldsymbol{E}_p^2 + \chi^{(2)}_{yxz} L_{yy}(\Omega) L_{xx}(\omega_1) L_{zz}(\omega_2) \boldsymbol{E}_p^2 + \chi^{(2)}_{yzx} L_{yy}(\Omega) L_{zz}(\omega_1) L_{xx}(\omega_2) \boldsymbol{E}_p^2 \quad (3)$$

where the effective nonlinear coefficient $\chi^{(2)}_{spp}$ includes contributions from multiple tensor elements weighted by Fresnel factors. Explicitly, it contains both achiral (e.g., $\chi^{(2)}_{yxx}$, $\chi^{(2)}_{yzz}$) and chiral (e.g., $\chi^{(2)}_{yxz}$, $\chi^{(2)}_{yzx}$) components.

The nonlinear susceptibility tensor elements in the laboratory coordinates $(x, y, z)$ is related to that in the crystal coordinates $(a, b, c)$ via:

$$\chi^{(2)}_{lmn} = \sum_{ijk} \chi^{(2)}_{ijk} (\hat{l} \cdot \hat{i})(\hat{m} \cdot \hat{j})(\hat{n} \cdot \hat{k}) \quad (4)$$

with $l, m, n \subset \{x, y, z\}$ and $i, j, k \subset \{a, b, c\}$.

To isolate chiral contributions, we perform an azimuthal average over a full rotation around the surface normal (z axis):

$$\langle \chi^{(2)} \rangle_\varphi = \frac{1}{2\pi} \int_0^{2\pi} \chi^{(2)}(\varphi) \, d\varphi \quad (5)$$

For a z-cut (001) surface, the crystal axes relate to the laboratory coordinates via

$$\hat{\boldsymbol{a}} = \cos\varphi\, \hat{\boldsymbol{x}} + \sin\varphi\, \hat{\boldsymbol{y}}, \quad \hat{\boldsymbol{b}} = -\sin\varphi\, \hat{\boldsymbol{x}} + \cos\varphi\, \hat{\boldsymbol{y}}, \quad \hat{\boldsymbol{c}} = \hat{\boldsymbol{z}} \quad (6)$$

Substituting Eqs. (3–6) and averaging over $\varphi$ yields an effective response

$$\langle \chi^{(2)}_{spp} \rangle_\varphi = \frac{L_{yy}(\Omega)}{2} \left[ (\chi^{(2)}_{bac} - \chi^{(2)}_{abc}) L_{xx}(\omega_1) L_{zz}(\omega_2) + (\chi^{(2)}_{bca} - \chi^{(2)}_{acb}) L_{zz}(\omega_1) L_{xx}(\omega_2) \right] \quad (7)$$

which depends only on antisymmetric combinations of the susceptibility tensor elements.

If a horizontal mirror plane $\sigma_h$ is present (perpendicular to the $c$ axis), all components of the second-order nonlinear tensor $\chi^{(2)}_{ijk}$ containing an odd number of indices along $c$ are forbidden by symmetry. As a result, the azimuthally averaged response vanishes, $\langle \chi^{(2)}_{spp} \rangle_\varphi = 0$. For a vertical ($\sigma_v$) or diagonal ($\sigma_d$) mirror planes, which necessarily contains the *c* axis, the same conclusion can be reached by transforming to a rotated in-plane coordinate system (*a'*, *b'*, *c*), where *a'* is normal to the mirror plane and *b'* axis lies within it. In this frame, symmetry allows only tensor elements of the form $\chi^{(2)}_{a'a'c}$ and $\chi^{(2)}_{b'b'c}$. Expressing the crystallographic axes (*a*, *b*) in terms of (*a'*, *b'*) via a rotation by an angle $\theta$,

$$\hat{a} = \cos\theta\, \hat{a}' + \sin\theta\, \hat{b}', \quad \hat{b} = -\sin\theta\, \hat{a}' + \cos\theta\, \hat{b}' \tag{8}$$

and substituting into Eq. (7), one again finds that all antisymmetric contributions cancel upon azimuthal averaging, yielding $\langle \chi^{(2)}_{spp} \rangle_\varphi = 0$.

These results establish a general symmetry constraint: the presence of any mirror symmetry, $\sigma_h$, $\sigma_v$, or $\sigma_d$, eliminates the antisymmetric components of $\chi^{(2)}$ enforcing a zero azimuthally averaged response. Conversely, a nonzero $\langle \chi^{(2)}_{spp} \rangle_\varphi$ requires the absence of all mirror symmetries. Because the symmetry of phonons are determined by the crystal point group, any phonon mode that preserves a mirror operation yields a vanishing averaged response. Therefore, $\langle \chi^{(2)}_{spp} \rangle_\varphi$ provides a direct, symmetry-based criterion for identifying chiral phonons, selecting modes that are odd under mirror reflection and thus intrinsically chiral.

**Circularly Polarized Terahertz Difference Frequency Spectroscopy**

Circularly polarized terahertz difference-frequency spectroscopy was implemented using intrapulse difference-frequency generation from a broadband femtosecond laser pulse centered at 1030 nm, with a spectral bandwidth of approximately 80 nm. A 1-mm-thick quartz plate was used as a dispersive waveplate, providing a wavelength-dependent retardance varying continuously from 0.75 to 0.25 across the 1006 - 1064 nm spectral range (Fig. S3 in SI). As a result, the high- and low-frequency components within the same pulse acquire opposite circular polarizations, enabling circularly polarized difference-frequency excitation within a single optical beam.

In this configuration, the helicity difference between the two frequency components determines the pseudo-angular momentum (PAM) transferred from the optical field to the lattice, thereby selecting phonon modes according to symmetry-imposed selection rules. Importantly, this transfer of PAM does not by itself guarantee real-space angular momentum (AM), which must be established independently through the emitted terahertz field.

The temporal walk-off between the two polarization components, arising from the birefringence between the ordinary and extraordinary axes of 1-mm-thick quartz, is estimated to be less than 30 fs. This is significantly shorter than the period of the emitted terahertz radiation (~300 fs), ensuring strong temporal overlap of the excitation pathways. Consequently, the phonons are coherently driven by the combined circularly polarized field, rather than by two independent linearly polarized excitations, which is essential for generating well-defined angular-momentum-carrying lattice motion.

**Samples**

Left-handed $\alpha$-quartz, right-handed $\alpha$-quartz, and $\alpha$-TeO$_2$ single crystals were commercially purchased from HF-Kejing. All samples were z-cut and double-side polished.

# Extended data

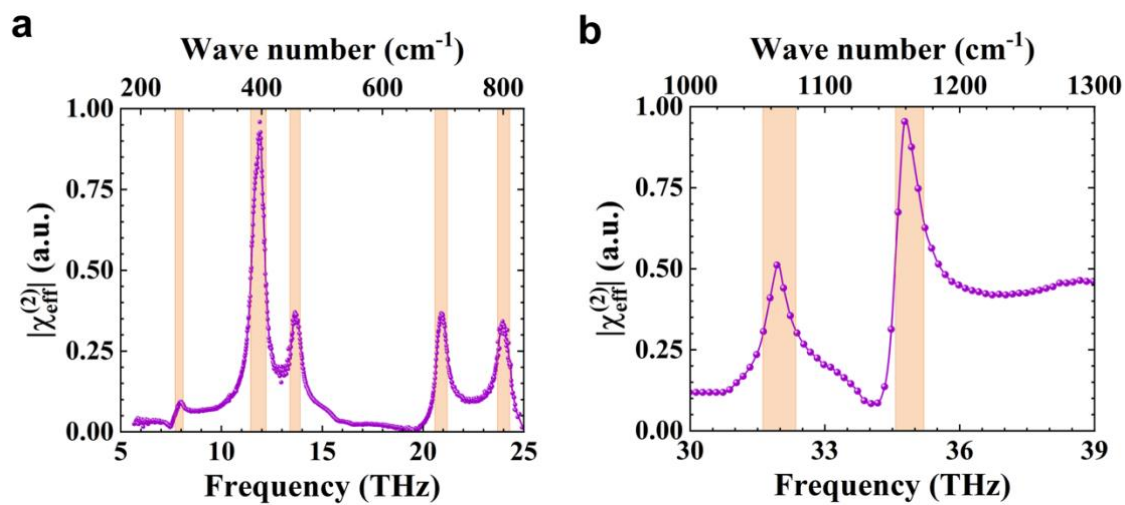

**Extended Data Fig. 1: Second-order nonlinear phonon spectra of *α*-quartz**. **a**, Terahertz sum-frequency spectrum of *α*-quartz (see Methods). **b**, Mid-infrared sum-frequency spectrum of *α*-quartz, reproduced from the Shen's group[35].

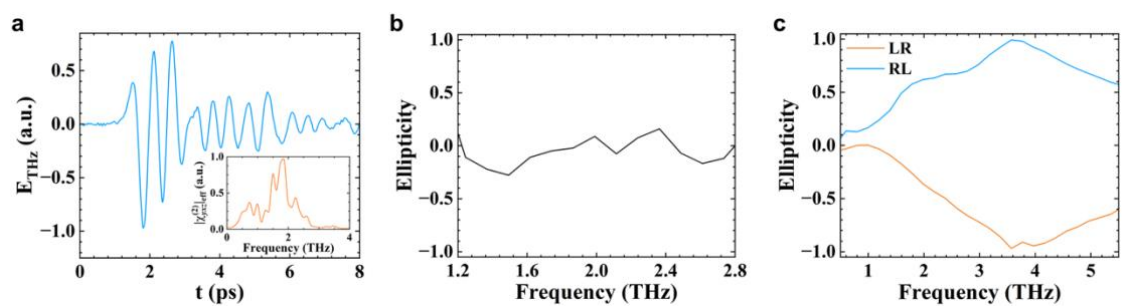

**Extended Data Fig. 2: Chirality without circular lattice motion: sucrose as a control system.** **a**, Time domain and frequency domain (insert) THz-DFS spectrum showing resonances near 1.5 THz and 1.8 THz. **b**, Ellipticity of terahertz field in frequency domain of sucrose crystal, defined by normalized third Stokes parameter[36]. **b**, Ellipticity of terahertz field in frequency domain of *α*-TeO2 crystal, for comparation.

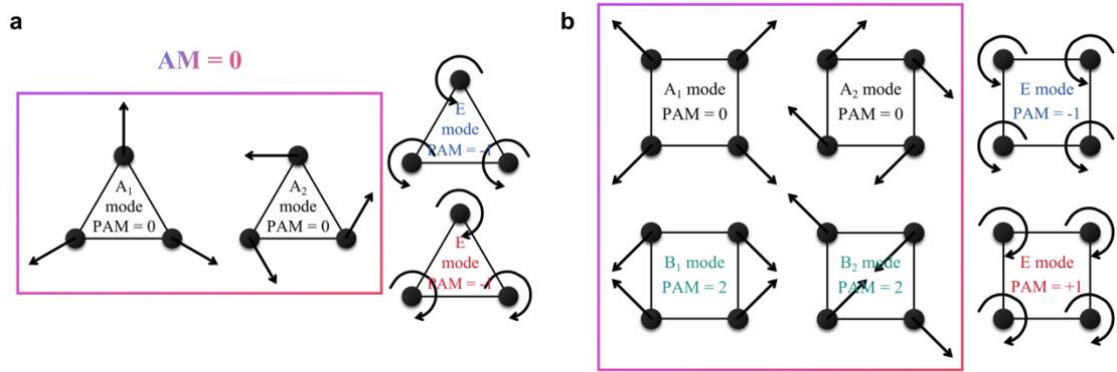

**Extended Data Fig. 3: Schematic illustration of motion for different modes.** a-b, Schematic illustration of motion for different modes in $C_3$ (a) and $C_4$ (b) crystal, the highlighted area is AM = 0 mode.

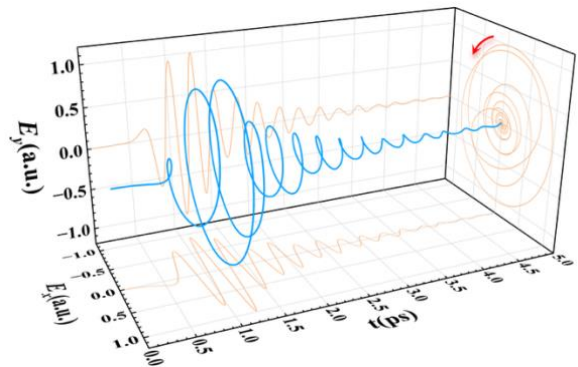

**Extended Data Fig. 4: Circularly polarized terahertz difference-frequency spectroscopy of α-TeO₂.** Circularly polarized terahertz spectra of α-quartz excited with left-handed pump with right-handed Stokes.